%% file: main.tex
\documentclass[sigplan,screen,nonacm]{acmart}
\usepackage{asystem-tech-report}
\usepackage{xspace}
\usepackage{xurl}
\usepackage{placeins}
\input{names}
\hypersetup{
  pdftitle={AInfer-PD: Communication-Safe In-Place Prefill-Decode Multiplexing for Distributed MoE Rollouts},
  pdfsubject={Technical report},
  pdfauthor={Guowei Wang, Chaokun Yang, Zhenxuan Pan, Yipeng Wei, Yuhong Guo, Minghua Zhu, Zechuan Zhang, Shuo Wan, Xiaowei Zhu},
  colorlinks=true,
  linkcolor=blue,
  citecolor=blue,
  urlcolor=blue
}

\title[AInfer-PD: Communication-Safe In-Place Prefill--Decode Multiplexing for Distributed MoE Rollouts]{%
\system: Communication-Safe\texorpdfstring{\\}{ }
In-Place Prefill--Decode Multiplexing\texorpdfstring{\\}{ }
for Distributed MoE Rollouts}
\reportauthor{Guowei Wang}{}
\reportauthor{Chaokun Yang}{}
\reportauthor{Zhenxuan Pan}{}
\reportauthor{Yipeng Wei}{}
\reportauthor{Yuhong Guo}{}
\reportauthorbreak
\reportauthor{Minghua Zhu}{}
\reportauthor{Zechuan Zhang}{}
\reportauthor{Shuo Wan}{}
\reportauthor{Xiaowei Zhu}{}
\reportaffiliation{}{Ant Group, China}

\begin{document}
\begin{abstract}
In agentic reinforcement learning (RL), asynchronous multi-turn trajectories
repeatedly introduce continuation prefills (P) while other trajectories remain
in decode (D).  Efficient rollout completion therefore requires D to advance
during long P executions.

In-place multiplexing shares devices, model weights, and KV state, but
phase-specific communication in distributed MoE introduces two constraints:
cross-rank collective ordering and ownership of mutable communication state.
Whole-prefill coordination can ensure safe ordering but delays D.  DeepEP's
normal and low-latency paths suit large P and small D batches, respectively,
but safely overlapping them requires separating their mutable communication state.

We present \system, which extends in-place P/D multiplexing to distributed MoE
rollouts.  \system establishes rank-consistent collective ordering at safe
communication boundaries, allowing successive D iterations to advance without
waiting for an entire prefill execution to be enqueued or completed.  It gives
P and D separate communication buffers and protocol state within one DeepEP runtime, allowing P to use
the normal path while D retains the low-latency path.  Together, these
mechanisms preserve useful P/D overlap while sharing model weights and KV
storage on the same devices.
Across repeated single-node prefill-intensive workloads, \system reduces
fixed-workload rollout completion time by 7.1--22.5\% relative to the same
\baseengine engine with P/D multiplexing disabled and by 24.8--32.9\% relative to
SGLang.  On two nodes, the reductions are 18.0--35.3\% and 18.3--31.8\%,
respectively.  In a same-engine ablation, fine-grained boundaries reduce
completion time by a further 8.6--19.8\% over whole-prefill asynchronous
enqueue.
\end{abstract}
\maketitle
\hypersetup{
  pdftitle={AInfer-PD: Communication-Safe In-Place Prefill-Decode Multiplexing for Distributed MoE Rollouts},
  pdfauthor={Guowei Wang, Chaokun Yang, Zhenxuan Pan, Yipeng Wei, Yuhong Guo, Minghua Zhu, Zechuan Zhang, Shuo Wan, Xiaowei Zhu},
  pdfsubject={Technical report}
}

\input{sections/01_introduction}
\input{sections/02_background_motivation}
\input{sections/03_overview}
\input{sections/04_pd_control_plane}
\input{sections/05_deepep_runtime}
\input{sections/06_implementation}
\input{sections/07_evaluation}
\input{sections/08_related_work}
\input{sections/09_discussion}
\input{sections/10_conclusion}

\raggedbottom
\appendix
\setcounter{figure}{0}
\setcounter{table}{0}
\renewcommand{\thefigure}{S\arabic{figure}}
\renewcommand{\thetable}{S\arabic{table}}
\input{supplement/a_protocol}
\input{supplement/b_lifecycle}
\input{supplement/c_waits}
\input{supplement/d_diagnostics}
\input{supplement/e_segments}
\input{supplement/f_live_rl}
\FloatBarrier
\bibliographystyle{ACM-Reference-Format}
\bibliography{references}
\end{document}

%% file: names.tex
\newcommand{\enginename}{AInfer}
\newcommand{\systemname}{AInfer-PD}
\newcommand{\baseengine}{\enginename\xspace}
\newcommand{\system}{\textsc{\systemname}\xspace}
\graphicspath{{figures/}}

%% file: sections/01_introduction.tex
\section{Introduction}

Rollout generation---the inference stage that produces trajectories for
optimization---often accounts for a large fraction of reinforcement-learning
(RL) training time~\cite{sheng2025hybridflow,fu2025areal,hu2026dora}.
In agentic RL, an agent decodes an action, waits for a tool or environment
response, appends the observation, and prefills the expanded context before
continuing generation.  Because trajectories differ in tool
latency, generation length, and termination time, continuation prefills recur
while other trajectories are still decoding.

The systems objective is therefore to finish a fixed collection of dependent
trajectories quickly.  Delaying decode postpones the environment interaction
that releases the next turn; delaying a returned prefill postpones that turn's
decode.  When the phases share accelerators, long and variable prefill (P)
batches also compete with latency-sensitive decode (D) for compute and
communication, increasing rollout completion time.  Efficient execution therefore requires D to keep
advancing during long P executions.

\begin{figure}[t]
  \centering
  \includegraphics[width=\columnwidth,trim={0 11.5bp 0 5.5bp},clip]{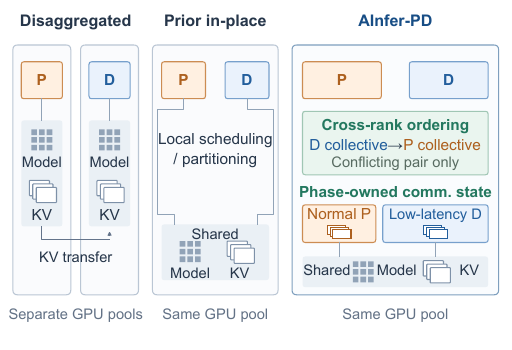}
  \caption{P/D execution choices.  Instance-level disaggregation separates model instances
  and transfers KV state.  In-place shares model and KV state.  \system
  adds cross-rank collective ordering and phase-owned expert-communication
  state to the in-place design.}
  \label{fig:intro-overview}
\end{figure}

Two broad approaches address this interference.  Instance-level P/D
disaggregation uses separate worker pools, duplicating model weights across
the pools and requiring KV transfer~\cite{patel2024splitwise,zhong2024distserve,qin2025mooncake}; in
multi-turn rollouts, this handoff can recur every turn.
In-place multiplexing shares devices, weights, and KV state, and limits local
interference through scheduling or resource
partitioning~\cite{agrawal2024sarathi,feng2025windserve,chen2026muxwise,
lin2026bullet,lee2026layeredprefill}.
Figure~\ref{fig:intro-overview} contrasts these execution choices.

Large MoE deployments combine attention data parallelism (DP), optional tensor
parallelism (TP) within each attention replica, and expert communication across a
wider group~\cite{deepseek2024v3}.  Phase-specific communication introduces
two constraints on concurrent P/D execution: cross-rank collective ordering
and ownership of mutable communication state.

The ordering constraint arises when P and D use intersecting communication
groups.  In common serving paths, P uses a model-wide TP AllReduce while
graph-replayed D uses a DP-attention
ReduceScatter/AllGather.  Different ranks can issue the P and D collectives in
opposite orders, leaving each collective waiting for ranks occupied by the
other.
Coordinating P and D at whole-prefill boundaries can ensure consistent
ordering, but defers D until an entire prefill execution has been enqueued or has
completed on the GPU.
The challenge is to establish the required order without making D wait for
the whole prefill execution.

The ownership constraint arises in DeepEP's phase-specific expert
communication~\cite{deepseek2025deepep}.  Its normal path suits large P batches,
while its low-latency path suits small D batches.  The two paths share mutable
communication state, including buffers and counters, which limits concurrent
use and forces D onto the normal
path in the shared-state P/D execution.  Preserving both specialized paths
therefore requires defining which communication state each phase must own
within the shared runtime.

\begin{figure}[t]
  \centering
  \includegraphics[width=\columnwidth,trim={0 10bp 0 5bp},clip]{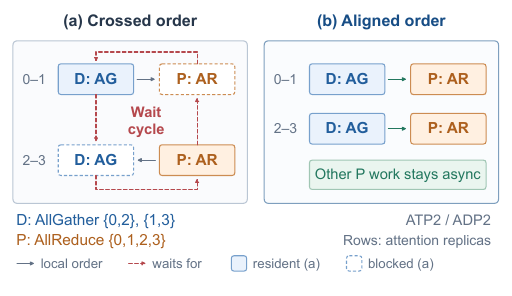}
  \caption{A crossed collective-order failure in an ATP2/ADP2 path.
  (a) Opposite orders of P full-group AllReduce and D DP-attention AllGather
  form a device progress cycle.  (b) \system orders the conflicting pair
  D-before-P on every rank; other P computation and nonintersecting
  communication remain asynchronous.}
  \label{fig:crossed-topology}
\end{figure}

\begin{figure*}[!t]
  \centering
  \includegraphics[width=\textwidth,trim={0 0.5bp 0 0bp},clip]{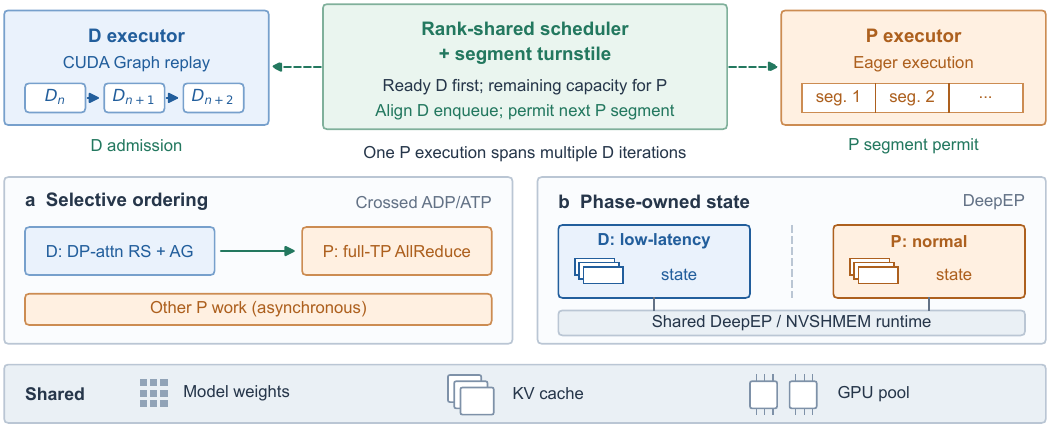}
  \caption{\system execution and communication paths.  The rank-shared
  scheduler and segment turnstile interleave D iterations with segments of
  one P execution.  (a) Selective ordering places the conflicting P full-TP
  collective after D's DP-attention collectives.  (b) D and P independently
  own low-latency and normal-path DeepEP state.  Both phases share model
  weights, KV cache, and GPUs; DeepEP retains one process-level runtime.}
  \label{fig:system-overview}
\end{figure*}

We present \system for communication-safe in-place P/D multiplexing in
distributed MoE rollout engines.  This paper makes three contributions:

\begin{itemize}
  \item We design a rank-aligned segment turnstile and selective device
  ordering for conflicting P/D collectives, allowing D to advance between
  segments of a long prefill execution.
  \item We extend DeepEP with phase-owned protocol and transport state,
  enabling normal-P and low-latency-D communication to coexist within one
  runtime while weights and KV storage remain shared.
  \item We integrate the mechanisms into an MoE rollout engine and evaluate
  communication safety, waiting induced by coordination, and rollout benefits
  across parallel topologies, precisions, multi-token prediction (MTP) modes, rollout pressure, and
  one- and two-node deployments.
\end{itemize}

We evaluate \system using anonymized internal RL traces on one- and two-node
H20-3E deployments.  Across the
repeated single-node prefill-intensive
profiles in Figure~\ref{fig:fresh-e2e}, \system reduces fixed-workload rollout
completion time by 7.1--22.5\% relative to the same \baseengine engine with P/D
multiplexing disabled and by 24.8--32.9\% relative to SGLang.  A two-node
campaign yields reductions of 18.0--35.3\% and 18.3--31.8\%, respectively,
with a TTFT trade-off.  Table~\ref{tab:ordering-ladder} reports a same-engine
ablation in which segment-level coordination improves completion by 8.6--19.8\%
over whole-prefill asynchronous enqueue.

%% file: sections/02_background_motivation.tex
\section{Background and Motivation}
\label{sec:motivation}

\subsection{Rollout Execution and P/D Scheduling}

We treat the full set of dependent trajectories as the unit of rollout work.
The completion-relative client wait is the interval between completion of one
turn's generation and release of its dependent continuation prefill.  It
includes environment latency and client-side response handling.  Short waits
increase recurrent P pressure; long waits leave more D-only execution.

Within a shared worker, chunked prefill mitigates decode interference, but its
chunk size or per-step token budget trades prefill efficiency against decode
iteration latency: smaller chunks provide more frequent opportunities for D
to advance, whereas larger chunks can improve P efficiency~\cite{agrawal2024sarathi}.
Chunked prefill can support attention data parallelism with MoE expert
parallelism by coordinating P chunks and D work within a common distributed
iteration.  When D advances only at these iteration boundaries, its progress
remains coupled to the P work admitted across participating replicas.
Changing chunk size still requires consistent collective participation and
communication-path selection across ranks.

\subsection{Parallel Organization and Phase-Specific Paths}

Large MoE inference commonly combines attention data parallelism (DP), tensor
parallelism (TP) within an attention replica, and expert communication across
a wider group~\cite{deepseek2024v3}.  Each attention-DP replica owns disjoint
sequences and KV state and schedules its own ready work.  The replicas can
therefore enter P and D at different times even though model or expert
communication crosses replica boundaries.

For world size $g$, let $d$ and $e$ denote the attention data-parallel (ADP)
and expert-parallel (EP) widths.  The corresponding attention and expert
tensor-parallel widths are $g/d$ (ATP) and $g/e$ (ETP).  Unless stated
otherwise, ETP is one; we therefore abbreviate a deployment by its nontrivial
EP and ADP dimensions.  For example, EP8/ADP2 denotes eight-way EP with two
attention-DP replicas and ATP4.

In practical MoE serving implementations, P and D may follow different
collective paths.  This mapping is not required by ADP: it lets variable-length
P use an eager model-wide TP AllReduce while padded D replays a DP-attention
ReduceScatter/AllGather.  Both SGLang~\cite{zheng2024sglang} and \baseengine use
this mapping in paths that combine ADP and ATP with either MoE TP or the
original EP backend.

DeepEP provides specialized paths for expert
parallelism~\cite{deepseek2025deepep}.  Its normal mode targets
throughput-oriented dispatch for large P batches, while its low-latency mode
targets latency-sensitive dispatch for small D batches.

\subsection{Communication Constraints on P/D Coexistence}

\paragraph{Cross-rank ordering.}
The P and D collective groups described above intersect.  Independently
scheduled attention replicas can enqueue their collectives in different
cross-rank orders, creating the distributed progress cycle in
Figure~\ref{fig:crossed-topology}.  When multiple communicators share a device,
NCCL requires their collectives to be launched in a consistent order across
ranks; CUDA Graph launches follow the same rule~\cite{nvidia2026ncclorder}.
Separate P/D communicators alone do not establish this cross-communicator
order.  Whole-prefill coordination can establish the order but couples D's
progress to the entire P enqueue or completion interval.

\paragraph{Mutable-state ownership.}
DeepEP's normal and low-latency modes were designed as alternative execution
modes rather than concurrent tenants: their mutable protocol and transport
state is shared.  In the shared-state execution, P/D coexistence therefore
falls back to the normal mode, making D communication slower than on the
low-latency path.  Concurrent use of the specialized paths requires separate
ownership of the state modified by each phase, even while process-level
initialization and physical resources remain shared.

%% file: sections/03_overview.tex
\section{Design Overview}
\label{sec:overview}

As Figure~\ref{fig:system-overview} shows, \system shares model weights, KV
cache, and devices between P and D.  A rank-shared scheduler controls phase
admission.  Communication coordination limits how much P work D must wait for,
while phase-owned DeepEP state preserves normal-P and low-latency-D execution
within one runtime.

\subsection{In-Place P/D Execution}

At each engine iteration, a rank-shared planner admits ready D first and uses
the remaining batch and KV capacity for P.

A single P execution typically spans the duration of several D iterations.
The backend exposes points where P can pause further communication submission
while already-enqueued GPU work continues.  These safe boundaries delimit P
segments.  A rank-aligned protocol, called the segment turnstile, coordinates
each round: all ranks first enqueue one D iteration in a consistent order,
then permit P to submit the next segment.  Repeating the round interleaves a
long P execution with successive D iterations; the segment policy chooses how
many boundary intervals each round admits.

\subsection{Ordering ADP/ATP Collectives}

For the intersecting ADP/ATP collectives described in
Section~\ref{sec:motivation}, the distributed turnstile establishes the same
host-enqueue order on every participating rank.  The conflicting P AllReduce
is submitted on D's graph-launch stream, so stream order places it after D.
Other P computation and nonintersecting communication remain on their
background streams.  This selective routing supplies the corresponding device
order without serializing all P/D work.
Section~\ref{sec:control-plane} defines the protocol and its progress argument.

\subsection{Concurrent DeepEP Communication}

\system assigns each phase its own mutable DeepEP protocol and transport
state while retaining one process-level runtime.  P uses the normal path, and
D keeps the low-latency path while P is active.  To connect this execution to
the turnstile, the normal dispatch is split so P can pause before submitting
routed data movement.  P waits for turnstile permission to resume submission,
while D proceeds on its own low-latency state.
Section~\ref{sec:deepep} details this design.

%% file: sections/04_pd_control_plane.tex
\section{Ordering Phase-Specific Collectives}
\label{sec:control-plane}

\subsection{Ordering Requirements}

In Figure~\ref{fig:crossed-topology}'s ATP2/ADP2 case,
graph-replayed D reaches a DP-attention AllGather over peers \{0,2\} (and
symmetrically \{1,3\}), while P reaches a full-group AllReduce over
\{0,1,2,3\}.  If different ranks make different kernels resident first, each
can wait for peers occupied by the other, closing a global cycle despite valid
local streams.

Avoiding this outcome requires rank-consistent host enqueue and a compatible
device order for the conflicting kernels.  \system builds this order in three
steps: ranks expose communication-safe P boundaries, a distributed turnstile
uses each boundary to establish a common host-enqueue phase for D, and a
selective stream route realizes that order for the conflicting device kernels.

\subsection{Rank-Aligned Segment Boundaries}

Each P execution forms an epoch $e$ with an ordered sequence of safe boundaries.
Exactly one backend publishes these boundaries: the non-DeepEP path at
registered layer boundaries, or DeepEP after notification in split-phase
normal dispatch.  The segment span specifies how many boundary intervals P
may submit before the next D iteration.  At startup, ready D is enqueued
before P enters CUDA.

\begin{figure}[t]
  \centering
  \includegraphics[width=\columnwidth,trim={0 5bp 0 5bp},clip]{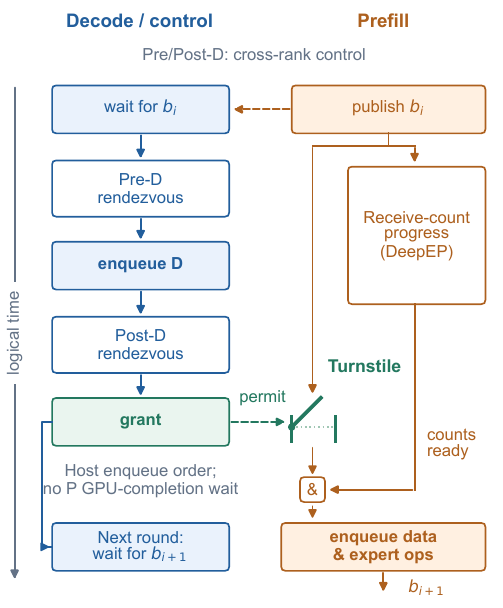}
  \caption{Rank-aligned segment turnstile.  $b_i$ and $b_{i+1}$ are
  policy-selected endpoints and may span multiple backend boundaries.  D
  waits for $b_i$, aligns ranks before and after enqueue, then grants the
  next P segment.  DeepEP data submission requires both receive counts and
  the permit; the next D round waits for $b_{i+1}$.}
  \label{fig:turnstile}
\end{figure}

\subsection{Distributed Segment Turnstile}

Safe boundaries define where D may enter; the turnstile coordinates when that
insertion occurs across ranks (Figure~\ref{fig:turnstile}).  P pauses submission
at the permitted endpoint.  The pre-D CPU rendezvous ensures that every rank
has reached that endpoint or finished its P epoch before D enqueue.  The
post-D rendezvous ensures that every rank has enqueued D before any rank
releases the next P segment.

Idle ranks issue matched no-work calls, and ranks whose P epoch has completed
continue through the matched control phases.  Neither CPU rendezvous enqueues
a GPU collective or waits for GPU completion; already-submitted P work continues.
The full controller state machine, including initialization and validation
checks, is specified in \hyperref[app:protocol]{Appendix~\ref*{app:protocol}}.

\subsection{Selective Device Ordering and Progress}

Host agreement alone does not order independent streams.  At each crossed-path
boundary, \system places the registered P full-TP AllReduce on D's graph-launch
stream.  Stream FIFO orders D before the following P collective locally, and
the turnstile aligns this order across ranks.  The pair does not overlap, but
other P work remains on its background route.  Repeating the order lets D
advance throughout P; paths without the registered collective keep their
original stream route.

Consider an interior turnstile step $j$ in epoch $e$.  Let $G_{e,j}^r$ be
rank $r$'s D graph launch, and let $R_{e,j}^{r,-}$ and $R_{e,j}^{r,+}$ be the
routed P calls immediately before the endpoint wait and after the next permit.
Let $\prec_r$ denote enqueue order on the shared stream and therefore execution
order under CUDA stream FIFO.  For every participating rank $r$, the protocol
establishes
\[
  R_{e,j}^{r,-}\prec_r G_{e,j}^r\prec_r R_{e,j}^{r,+}
\]
The initial endpoint has no preceding P call, and a completed epoch has no
following P call; each requires only the applicable half-order.

\paragraph{Why the cycle is removed.}
The turnstile aligns the enqueue phases across ranks, and stream FIFO
enforces the resulting order locally~\cite{nvidia2026cudastreamorder,nvidia2026cudagraph}.
Figure~\ref{fig:crossed-topology}'s cycle requires opposite cross-rank orders,
so it cannot close.  The argument covers the registered collective pair under
matched participation and backend progress.

%% file: sections/05_deepep_runtime.tex
\section{DeepEP Communication-State Isolation}
\label{sec:deepep}

To preserve concurrent normal-P and low-latency-D execution, \system gives
each phase exclusive ownership of its mutable DeepEP communication state.
The design isolates protocol objects and transport state within one runtime,
then exposes a normal-dispatch boundary for the segment turnstile.

\subsection{Phase-Owned State}

Objects modified by communication kernels belong exclusively to P or D.
Table~\ref{tab:deepep-ownership} lists the shared resources and phase-owned
state, including queue-pair (QP) ranges.

\begin{table}[t]
  \caption{Ownership in the concurrent DeepEP runtime.}
  \label{tab:deepep-ownership}
  \centering
  \small
  \begin{tabular}{@{}p{0.22\columnwidth}p{0.70\columnwidth}@{}}
    \toprule
    Ownership & Objects \\
    \midrule
    Shared & NVSHMEM initialization, rank topology, physical links, model
      weights, KV storage \\
    P-owned & Normal buffer, counters, workspace, stream, events, QP range \\
    D-owned & Low-latency buffer, counters, workspace, stream, events,
      double-buffer index, QP range \\
    \bottomrule
  \end{tabular}
\end{table}

Phase identity selects state explicitly.  D retains stable CUDA Graph
addresses, while P uses separate eager inputs and cannot reset D metadata.
NIC bandwidth, caches, and unreserved SM capacity remain shared.

\subsection{Network and Lifetime Isolation}

Both phase-owned buffers use one global-rank NVSHMEM space but register
disjoint per-peer QP intervals, preventing one phase from advancing
the other's transport state.  Construction validates peer mappings before
communication, and buffer-local synchronization avoids cross-phase fences.
Shared initialization remains live while either phase or a captured graph can
access it.  Separate communication-SM budgets bound cooperative kernels; their
configured footprint is reported in Section~\ref{sec:evaluation}.

\subsection{Exposing a Safe Prefill Boundary}

With state ownership established, the normal P path supplies the safe boundary
used by the turnstile in Section~\ref{sec:control-plane}.  Normal DeepEP dispatch
combines expert notification, receive-count exchange,
tensor allocation, and routed data movement.  \system splits it after
notification, when peers know the dispatch shape but before P submits data
movement.  P then obtains receive counts in the background and submits data
only after both counts and turnstile permission are available; D proceeds on
its low-latency state.  Figure~\ref{fig:turnstile} shows these dependencies.

The pending P dispatch retains the tensors and stream state needed to finish;
its host-side count wait does not block D.  At most one dispatch is pending per
P buffer, and explicit transitions prevent cancellation from reusing partially
advanced state.

Because device-initiated communication cannot be retracted after notification,
an abandoned operation retires its buffer until a quiescent cleanup point.

%% file: sections/06_implementation.tex
\section{End-to-End Integration}
\label{sec:implementation}

To integrate the ordering and ownership mechanisms, the engine carries phase
identity from scheduling into attention and expert communication.  This
identity selects the collective route and DeepEP state; it is not inferred
from a host thread or CUDA stream.

\subsection{Scheduling, KV, and Graph State}

At each scheduling step, the engine selects the sequences for D, prepares
their execution inputs, and reserves KV capacity for decode and any MTP
lookahead before admitting P into the remaining batch and KV capacity.
The rank-shared plan protects D's KV slots and prevents local ready queues from
selecting incompatible phase paths.

Each P request owns its KV reservation until the request is promoted to decode or cancelled;
submitted P work must drain before cancelled storage is reused
(\hyperref[app:lifecycle]{Appendix~\ref*{app:lifecycle}}).

D graph capture binds phase-owned inputs, metadata, stream routes, and stable
low-latency DeepEP state.  Background P uses disjoint eager inputs, so the same
D graph can be replayed whether P is active or idle.  MTP drafting and
verification form one D iteration and finish enqueue before the post-D
rendezvous releases P.

\subsection{Prefill Policy Interface}

Given the backend's ordered sequence of safe P boundaries, the segment policy
chooses how many boundary intervals to admit per round.  We
evaluate fixed spans and a load-selected policy that chooses among validated
backend- and precision-specific spans according to current D occupancy and P
pressure.  These policies trade D waiting time against control overhead while
retaining the same epoch checks, rendezvous order, and phase-state isolation.

%% file: sections/07_evaluation.tex
\section{Evaluation}
\label{sec:evaluation}

We first characterize recurrent prefill work and measure rollout completion
(RQ1--RQ2).  We then examine whether ordering removes the observed cycle and
how much D must wait under different coordination policies and segment sizes
(RQ3), followed by normal-P and low-latency-D coexistence in DeepEP (RQ4).
RQ5 varies rollout pressure; RQ6 evaluates cross-node and live-RL execution.

\begin{figure*}[!t]
  \centering
  \includegraphics[width=\textwidth,trim={0 1bp 0 0bp},clip]{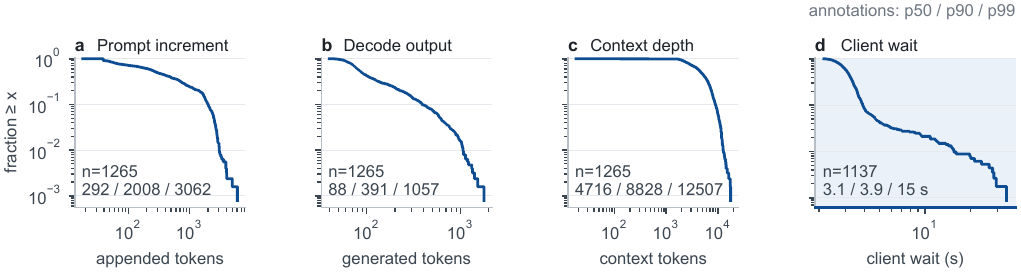}
  \caption{Distributions in the anonymized internal reinforcement-learning
  rollout trace.  Panels show empirical CCDFs of
  (a) newly appended prompt tokens, (b) generated tokens, (c) retained context
  length, and (d) trace-derived completion-relative client wait.  Annotations
  give sample count and p50/p90/p99.  Both axes are logarithmic.}
  \label{fig:workload-distribution}
\end{figure*}

\subsection{Methodology}

\textbf{Systems and hardware.}
The main campaign uses eight NVIDIA H20-3E GPUs.  Its software stack includes
PyTorch 2.8, CUDA 13.0, NCCL 2.27.7, DeepEP 1.2.1, and SGLang 0.5.15
~\cite{zheng2024sglang}.  BF16 and FP8 runs use
42-layer internal MoE checkpoints with hidden size 2,560, 512 routed experts,
and top-8 routing.  A two-node campaign uses 16 H20-3E GPUs, a larger internal
checkpoint, the same SGLang version, and a separate anonymized RL trace.  RQ3
uses eight H200 GPUs for the crossed-cycle profile and H20-3E for model-free
rank-skew stress; these diagnostic runs are excluded from performance results.

Table~\ref{tab:topology-map} lists the evaluated parallel configurations.
H1/H2 use tensor-parallel experts and state ETP explicitly; the remaining
profiles have ETP1.  The eight-GPU crossed profile EP8/ADP2 therefore has ATP4.

\begin{table}[!t]
  \caption{Evaluation profiles.  H1--H4 use BF16 and the non-DeepEP
  MoE backend; crossed
  cells span BF16/FP8 and MTP off/on.  ATP and ETP are derived from world size
  as $g/d$ and $g/e$.}
  \label{tab:topology-map}
  \centering
  \scriptsize
  \setlength{\tabcolsep}{2.2pt}
  \begin{tabular}{@{}lcccc@{}}
    \toprule
    Profile & Nodes$\times$GPUs & EP/ADP & ATP/ETP & Backend / MTP \\
    \midrule
    H1 & $1\times4$ & EP1/ADP2 & ATP2/ETP4 & non-DeepEP / off \\
    H2 & $1\times4$ & EP1/ADP2 & ATP2/ETP4 & non-DeepEP / on \\
    H3 & $1\times8$ & EP8/ADP8 & ATP1/ETP1 & non-DeepEP / off \\
    H4 & $1\times8$ & EP8/ADP8 & ATP1/ETP1 & non-DeepEP / on \\
    Crossed & $1\times8$ & EP8/ADP2 & ATP4/ETP1 & DeepEP / off,on \\
    Two-node & $2\times8$ & EP16/ADP8 & ATP2/ETP1 & DeepEP / off \\
    \bottomrule
  \end{tabular}
\end{table}

\textbf{Workload and protocol.}
The main replay preserves request order and sizes, multi-turn dependencies,
output lengths, and client waits from an anonymized internal RL trace.  For
continuation $i$, the driver applies the same configured wait $\delta_i$ after
the preceding turn completes.  Every main-campaign performance run finishes 1,265 requests
from 128 conversations, consuming 6,574,104 input and producing 220,862 output
tokens.  Headline cells use five repetitions and other single-node cells use
three; we report medians and interquartile ranges (IQRs).  All runs complete without eviction or
runtime failure.

\textbf{Baselines.}
We use \baseengine with strict prefill priority and P/D multiplexing
disabled as the same-engine baseline.  \system enables the complete design
on the same underlying engine and kernels.  SGLang provides an independent
whole-stack baseline.  SGLang uses the same requests, token budgets, caching, overlap
scheduling, and MTP setting; matched MTP uses three-step top-1 NextN with four
draft tokens.

\textbf{Metrics.}
Primary metrics are fixed-workload makespan and completed requests/s,
reciprocal views of 1,265 completions.  Request completion spans submission to
the final token; trajectory completion spans the first release to the last
turn's final token, including configured waits.  We also report p99 TTFT and
mean D wait.  Makespan is primary because training consumes the completed
rollout.

\textbf{Functional regression.}
We run a GSM8K exact-answer regression on \baseengine, \system, and SGLang in BF16
with MTP off/on, and on DeepEP-enabled \system in BF16/FP8 with MTP off/on.
All ten runs are valid, and paired BF16 accuracy differs from \baseengine by at most
2.20 points.

\subsection{RQ1: What Recurs on the Rollout Critical Path?}

The trace spans up to ten causally dependent turns and includes 1,137
continuations released after a preceding generation, so P work recurs throughout
the rollout.  Figure~\ref{fig:workload-distribution} shows long-tailed prompt
increments, generated outputs, and retained context; the latter reaches
12,507 tokens at p99.  Unequal turn lengths and asynchronous continuation
releases create repeated opportunities for phase skew across attention replicas.

Trace-derived client wait has a 3.10~s median and 15.30~s p99.  The headline
configuration releases continuations immediately; RQ5 scales the assigned
wait vector by 0, 0.01, 0.03, and 0.10 and reports the original vector
separately.

\subsection{RQ2: End-to-End Rollout Completion}

\begin{figure*}[!t]
  \centering
  \includegraphics[width=\textwidth,trim={0 1bp 0 3bp},clip]{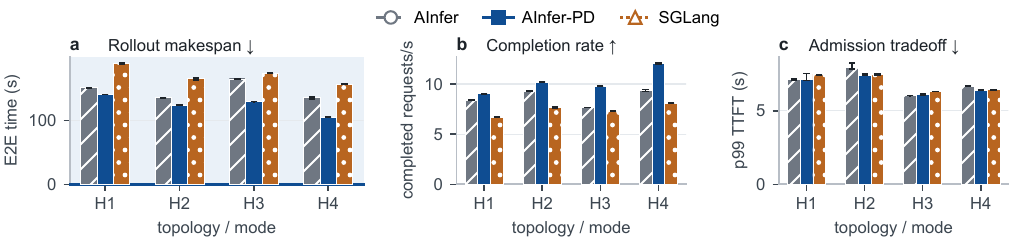}
  \caption{Fixed-workload end-to-end comparison using the profiles in
  Table~\ref{tab:topology-map}.  Panels report
  (a) replay makespan, (b) completed requests/s, and (c) p99 TTFT.  Bars are
  medians; error bars are IQRs.  Since every run completes
  1,265 requests, panels a and b are reciprocal views.}
  \label{fig:fresh-e2e}
\end{figure*}

Figure~\ref{fig:fresh-e2e} compares four high-P-pressure profiles with no added
continuation idle.  Across H1--H4, \system
reduces E2E time relative to \baseengine by 7.1\%, 8.7\%, 21.5\%, and
22.5\%.  Relative to SGLang, E2E falls by 24.8--32.9\%.  The result holds
across both topologies and with MTP on or off.

H1--H2 exercise the crossed ADP/ATP ordering path.  H3--H4 have ATP1 and need
neither the selective route nor DeepEP state isolation; they measure the
underlying benefit of in-place P/D scheduling.

Across H1--H4, p99 request completion falls by 21.3--37.9\% versus \baseengine and
39.3--44.0\% versus SGLang; trajectory completion falls by 8.2--22.7\% and
27.0--33.4\%, respectively.  P99 TTFT remains within $-6.3$ to $+1.2$\% of
\baseengine.

\subsection{RQ3: Collective Safety and Decode Waiting}

\textbf{Removing the observed cycle.}
Communication-group separation alone repeatedly stalls the crossed ATP2/ADP2
path.  Figure~\ref{fig:crossed-topology} illustrates the cycle observed at a
four-rank D AllGather/P AllReduce frontier at 418/1,265 completed requests
in an instrumented replay.  Table~\ref{tab:collective-repair} summarizes the
controls: reducing either topology axis or lowering kernel residency avoids
the stall, and selective ordering completes all three full replays in the
crossed topology.  Rank-skew stress completes 2,560/2,560
turnstile cycles (controls and stress settings in \hyperref[app:diagnostics]{Appendix~\ref*{app:diagnostics}}).

\textbf{Waiting under safe coordination.}
Having established safe collective ordering, we next ask whether D can
advance promptly under that ordering.  We compare three coordination policies
that differ in how much P progress D must wait for, while holding admission,
lifecycle management, phase-owned DeepEP state, model, and engine settings
fixed.  \emph{Global-Complete} uses a whole-P GPU fence;
\emph{Global-Enqueue} acknowledges all-rank whole-epoch host enqueue without
waiting for GPU completion; \emph{Fine-grained} advances at required segment
boundaries.

\begin{table}[t]
  \caption{Matched crossed-topology safe-PD coordination policies (BF16,
  DeepEP, MTP off).  Makespan and TTFT are median (IQR), $n=3$; mean D wait
  is the median across runs of the per-iteration mean.}
  \label{tab:ordering-ladder}
  \centering
  \footnotesize
  \setlength{\tabcolsep}{2.5pt}
  \begin{tabular}{@{}lrrr@{}}
    \toprule
    Policy & Mean D wait & Makespan & p99 TTFT \\
      & (ms) $\downarrow$ & (s) $\downarrow$ & (s) $\downarrow$ \\
    \midrule
    \multicolumn{4}{@{}l}{\textit{Trace wait}} \\
    Global-Complete & 138.5 & 179.79 (1.40) & 6.26 (.65) \\
    Global-Enqueue & 127.3 & 181.39 (.73) & 7.14 (.73) \\
    Fine-grained & 17.5 & 165.83 (1.81) & 6.37 (.81) \\
    \midrule
    \multicolumn{4}{@{}l}{\textit{No wait}} \\
    Global-Complete & 122.6 & 181.12 (1.79) & 6.64 (.54) \\
    Global-Enqueue & 114.1 & 168.88 (1.15) & 6.88 (.08) \\
    Fine-grained & 18.0 & 135.49 (2.17) & 7.34 (.56) \\
    \bottomrule
  \end{tabular}
\end{table}

Removing the whole-P completion fence does not remove D's wait for whole-P
submission.  Under Global-Enqueue, mean D wait remains 127.3~ms with trace
waits and 114.1~ms without waits (Table~\ref{tab:ordering-ladder}).
Fine-grained coordination reduces these waits to 17.5 and 18.0~ms,
respectively, and lowers makespan by 8.6\% and 19.8\%.  The improvement
therefore persists beyond asynchronous whole-prefill enqueue and under the
original trace-derived waits.  With no added wait, p99 TTFT rises from
6.88 to 7.34~s, reflecting the trade-off between rollout completion and
prefill latency.

All six Global-Enqueue runs acknowledge enqueue before P GPU completion in
some iterations, confirming that this baseline does not impose a whole-P
completion fence.

\textbf{Execution overlap.}
In a 40-step repaired-path diagnostic window, the registered NCCL pair has
zero overlap, while P computation overlaps D for 6.3--6.5\% on the two
observed ranks (\hyperref[app:diagnostic-traces]{Appendix~\ref*{app:diagnostic-traces}}).  Selective ordering therefore preserves some
computation concurrency even though the conflicting collectives themselves
do not overlap.

\textbf{Segment granularity.}
Larger segments reduce P's wait for permission but increase D's boundary wait.
In all six selector runs, the runtime chooses span 2, which is best for BF16
and within 0.8\% of the best FP8 choice among the tested fixed spans 1--4
(\hyperref[app:segments]{Appendix~\ref*{app:segments}}).

\subsection{RQ4: Can DeepEP's P/D Protocols Coexist?}

\begin{figure}[t]
  \centering
  \includegraphics[width=0.90\columnwidth,trim={0 1.5bp 0 0.5bp},clip]{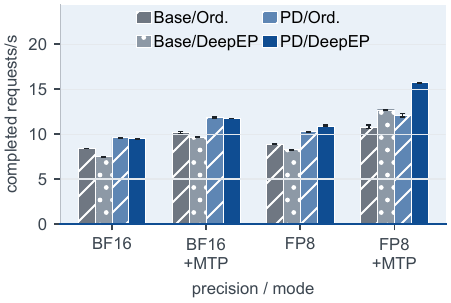}
  \caption{Backend interaction across precision and MTP settings in
  EP8/ADP2.  Base and PD denote \baseengine and \system; Ord. denotes the
  non-DeepEP backend.  Bars are medians; error bars are IQRs.}
  \label{fig:deepep-ablation}
\end{figure}

\textbf{Concurrent phase-specific paths.}
With phase-owned communication state, \system runs normal-P and low-latency-D
communication concurrently within one DeepEP runtime.  With DeepEP and FP8,
it reaches 9.18--14.74 requests/s in EP8/ADP8,
26.6--50.4\% above \baseengine and 36.0--62.8\% above SGLang.  The crossed
EP8/ADP2 configuration reaches 9.47--15.74 requests/s across precision and
MTP settings.

\textbf{Backend interaction.}
Figure~\ref{fig:deepep-ablation} compares the backends under the two
scheduling modes.  In
EP8/ADP2, \system improves request rate over \baseengine by 11.7--16.1\%
with the non-DeepEP backend and by 21.8--31.9\% with DeepEP.  Enabling DeepEP
changes request rate by $-11.3$ to $+17.8$\% for \baseengine and by $-1.7$ to
$+30.8$\% for \system; its effect therefore depends on the scheduling mode.
These comparisons measure the combined effect of backend choice and P/D execution.

\textbf{State and resource footprint.}
In EP8, DeepEP reserves 140.2~MiB per rank for normal-path NVLink payload and
325.0~MiB for low-latency RDMA payload: 465.2~MiB in both single- and dual-phase
configurations.  Concurrency adds P metadata and a 12-index interval in each
128-index per-peer QP namespace.  DeepEP's normal prefill path uses 12 of 132 SMs; NIC bandwidth
and residual SM capacity remain shared.

\subsection{RQ5: Sensitivity to Rollout Pressure}

We repeat the eight-GPU DeepEP topology for all three systems in BF16 with MTP
off.  P1--P4 multiply the trace-derived wait vector by 0, 0.01, 0.03, and 0.10
while preserving prompts and dependencies; the original vector is separate.
Table~\ref{tab:pressure-profiles} in
\hyperref[app:waits]{Appendix~\ref*{app:waits}} lists the resulting wait quantiles.

\begin{figure}[!t]
  \centering
  \includegraphics[width=\columnwidth,trim={0 0.5bp 0 3bp},clip]{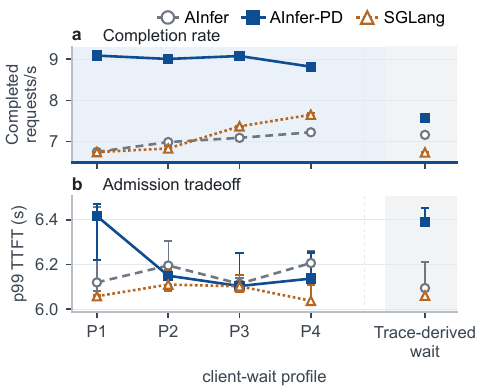}
  \caption{Sensitivity to completion-relative client wait.  P1--P4 scale
  the trace waits by 0, 0.01, 0.03, and 0.10; the shaded point uses the
  unmodified trace waits.  Panels show (a) completed
  requests/s and (b) p99 TTFT.  Points are medians; error bars are IQRs.}
  \label{fig:pressure-results}
\end{figure}

Figure~\ref{fig:pressure-results} shows the sensitivity to client wait.
Across P1--P4, \baseengine and SGLang reach 6.75--7.23 and 6.75--7.65 requests/s;
\system reaches 8.82--9.09, gains of 22.0--34.6\% and 15.2--34.7\%.
Its p99 TTFT is 6.10--6.42~s, at worst 4.9\% and 5.9\% higher.  P99 request
completion falls by 36.0--46.5\% and 28.8--45.2\%; trajectory completion by
18.6--26.9\% and 13.7--27.0\%.

With the original wait vector, \system's 166.93~s median makespan is 5.5\%
below \baseengine and 11.1\% below SGLang; p99 TTFT rises 4.9\% and 5.5\%.  Longer
client waits leave less overlap opportunity.

In a decode-only control, enabling the P/D path with the same kernels lowers
request rate by 6.7\%, motivating activation only when concurrent P work
exists.

\subsection{RQ6: Cross-Node and Live-RL Validation}

\begin{table*}[!t]
  \caption{Two-node, 16-H20-3E validation (BF16, FA3, MTP off,
  EP16/ADP8 (ATP2, ETP1), DeepEP, 32K-token P budget).  Each cell is a three-run mean;
  rows increase workload depth at matched within-row work.  SG denotes SGLang.}
  \label{tab:multinode}
  \centering
  \footnotesize
  \setlength{\tabcolsep}{2.0pt}
  \begin{tabular}{@{}rcrrrrrrrrrr@{}}
    \toprule
    & Context & \multicolumn{3}{c}{E2E (s)} &
      \multicolumn{3}{c}{Completed REQ/s} &
      \multicolumn{3}{c}{p99 TTFT (s)} \\
    \cmidrule(lr){2-2}\cmidrule(lr){3-5}\cmidrule(lr){6-8}\cmidrule(lr){9-11}
    Rounds & Avg/Max & SG & \baseengine & \system & SG & \baseengine & \system & SG & \baseengine & \system \\
    \midrule
    10  & 5,197/17,479   & 486.31   & 512.69   & 331.57   & 2.601 & 2.467 & 3.815 & 26.06 & 18.83 & 21.31 \\
    25  & 10,102/35,884  & 1,209.39 & 1,213.79 & 887.51   & 2.575 & 2.566 & 3.509 & 13.87 & 14.11 & 15.98 \\
    50  & 17,304/68,774  & 2,711.48 & 2,712.62 & 2,154.78 & 2.160 & 2.159 & 2.718 & 10.73 & 11.04 & 15.22 \\
    100 & 25,167/101,214 & 3,966.09 & 3,951.15 & 3,238.89 & 2.082 & 2.090 & 2.550 & 8.41  & 8.47  & 10.39 \\
    \bottomrule
  \end{tabular}
\end{table*}

Table~\ref{tab:multinode} holds within-row work identical: 1,265--8,258
requests, 6.57M--207.83M input tokens, and 0.22M--2.73M output tokens.  \system
cuts E2E 18.0--35.3\% versus \baseengine and 18.3--31.8\% versus SGLang; maximum
range/mean is 1.28\%, 1.20\%, and 2.09\% for SGLang, \baseengine, and \system,
respectively.  Across the four workload scales, relative to \baseengine, p99 TTFT
rises 13.2--37.9\%, exposing the completion-time/TTFT tradeoff under the
completion-oriented admission policy.  A more conservative P budget or segment
policy can prioritize TTFT at the expense of rollout completion.

Two independent live-RL runs compare \system with SGLang.  Summed rollout
time across steps 0--31 falls from 17.14 to 14.12 hours (17.6\%; per-step
results in \hyperref[app:live]{Appendix~\ref*{app:live}}).  Generated trajectories may diverge after each step,
so this is operational rather than paired evidence.

%% file: sections/08_related_work.tex
\section{Related Work}
\label{sec:related}

\paragraph{Agentic RL execution.}
HybridFlow, AReaL, and DORA improve placement, pipelining, and synchronization
between rollout, reward, and training stages in RL systems
\cite{sheng2025hybridflow,fu2025areal,hu2026dora}.  These systems primarily
address how stages share a cluster.  \system is complementary: it targets the
inference runtime within a rollout worker, where asynchronous multi-turn
continuations repeatedly introduce P work alongside active D.

\paragraph{P/D disaggregation.}
Splitwise, DistServe, and Mooncake place P and D on distinct workers or pools
and transfer KV state between them
\cite{patel2024splitwise,zhong2024distserve,qin2025mooncake}.  This placement
provides separate phase capacity and communication domains.  ExpertPlex uses a
hybrid organization that disaggregates attention while sharing a distributed
expert pool~\cite{wu2026expertplex}.  Its attention-initiated one-sided dispatch
and combine remove receiver-side progress dependencies that can otherwise
cause cross-phase deadlock.  \system instead operates within one
in-place model and KV replica, coordinating P/D collective order and separating
mutable communication state within that shared deployment.

\paragraph{In-place P/D multiplexing.}
Sarathi-Serve, Layered Prefill, and POD-Attention change prefill granularity or
joint execution to reduce decode contention
\cite{agrawal2024sarathi,lee2026layeredprefill,kamath2025podattention}.
WindServe, MuxWise, Bullet, semi-PD, Nexus, and DuetServe multiplex P and D within a
worker through stream scheduling, spatial partitioning, or phase-wise resource
management
\cite{feng2025windserve,chen2026muxwise,lin2026bullet,hong2025semipd,shi2025nexus,gao2026duetserve}.
NanoFlow overlaps compute-, memory-, and network-intensive nano-operations
through automatically generated pipelines, including tensor-parallel MoE
execution~\cite{zhu2025nanoflow}.  In distributed MoE
serving, attention replicas can issue intersecting P and D collectives in
different orders.  The cited designs do not specify rank-consistent ordering
for this case; \system coordinates the conflicting calls at safe boundaries
so D can advance within a long P execution.

\paragraph{Collective ordering.}
DFCCL prevents GPU collective deadlocks through daemon-kernel scheduling and
collective preemption~\cite{pan2025dfccl}.  NCCL implicit cross-communicator
ordering requires globally consistent host launches~\cite{nvidia2026ncclorder}.
\system coordinates conflicting calls at safe P/D segment boundaries while
leaving other P work asynchronous.

\paragraph{Distributed MoE communication.}
Lina prioritizes all-to-all during training and schedules inference resources
according to expert popularity~\cite{li2023lina}.
DeepEP provides normal and low-latency expert-communication
protocols~\cite{deepseek2025deepep}.  \system separates their mutable protocol
and transport state, enabling concurrent normal-P and low-latency-D execution
within one runtime.  This state-ownership requirement is distinct from
communication resource scheduling and cross-rank collective ordering.

%% file: sections/09_discussion.tex
\section{Discussion and Limitations}
\label{sec:discussion}

\paragraph{Applicability and ordering scope.}
Extending the
turnstile beyond the evaluated ADP/ATP paths to a new collective pair
requires identifying its participants, safe P boundary, device-order edge, and
no-work behavior.  The ordering result assumes healthy ranks and transport,
matched participation, and fair CUDA/NCCL progress.  Epoch validation detects
violations of this protocol rather than general system faults.

\paragraph{State ownership and shared resources.}
State isolation does not eliminate contention for shared NIC bandwidth,
caches, or unreserved SM capacity; admission and communication-SM budgets
regulate this contention.  The ownership principle may extend to other expert
runtimes with phase-specialized state, although we evaluate only DeepEP.  The
primary objective is rollout completion, and aggressive P admission can
increase TTFT.  Segment size trades P submission opportunities against D's
boundary wait without changing the ordering and ownership requirements;
online adaptation remains future work.

\paragraph{Evaluation scope.}
We evaluate one- and two-node H20-3E deployments, BF16/FP8, both MoE backends,
and MTP on/off; broader scales and hardware remain open.  Same-engine controls
isolate coordination policy with phase-owned DeepEP state held fixed;
backend comparisons evaluate phase-specific execution within the full system.
SGLang is a matched-work whole-stack comparison.
Live-RL trajectories can diverge and provide operational rather than paired
evidence.

%% file: sections/10_conclusion.tex
\section{Conclusion}

Multi-turn rollouts require decode to advance while other trajectories execute
long prefills.  \system enables this within a shared distributed MoE instance
through two mechanisms: coordination at safe communication boundaries avoids
making D wait for an entire P execution, and phase-owned DeepEP state lets
normal-P and low-latency-D paths coexist within one runtime.  Together, they
preserve useful P/D overlap while sharing model weights and KV storage.

Across the evaluated prefill-intensive single-node profiles, \system reduces
rollout makespan by 7.1--22.5\% versus the same engine with multiplexing
disabled and by 24.8--32.9\% versus SGLang; the two-node profiles yield
reductions of 18.0--35.3\% and 18.3--31.8\%, respectively.  With phase-owned
DeepEP state held fixed, segment-level coordination further reduces makespan
by 8.6--19.8\% over whole-epoch asynchronous enqueue in the measured crossed
topology.

%% file: supplement/a_protocol.tex
\section{Detailed Turnstile Protocol}
\label{app:protocol}

The controller identifies each P execution by an epoch $e$ and its safe
boundaries by monotonically increasing identifiers $i=0,1,\ldots$.
Exactly one backend publishes them: the non-DeepEP path at
registered layer boundaries, or DeepEP after notification in
split-phase normal dispatch.

The controller stores the largest reached boundary $r$ (initially $-1$),
permitted endpoint $p$, positive span $K$, and task state.  The endpoint
limits how far P may submit work before the next D iteration; $K$ is the
number of boundary intervals released after each D enqueue round.  Initially,
$p=0$ orders ready D before P enters CUDA; if D is already enqueued, P starts
with $p=K$.  The controller rejects stale or duplicate epochs,
missing producers, nonmonotone boundary publication, and endpoints beyond the
current permit.

\textsc{PreDRendezvous} and \textsc{PostDRendezvous} are all-rank CPU
control phases.  They carry no endpoint value and enqueue no GPU collective.
Before D enqueue, all ranks reach the permitted boundary or finish their P
epoch; after enqueue, all ranks acknowledge D submission before P is released.

\begin{center}
  \begin{minipage}{0.98\columnwidth}
  \footnotesize
  \textbf{Turnstile step for P epoch $e$ at permitted endpoint $p$}

  \setlength{\tabcolsep}{2.0pt}
  \begin{tabular}{@{}r p{0.85\columnwidth}@{}}
    \toprule
    & Action \\
    \midrule
    P1 & If $p=0$, wait before CUDA.  Publish $(e,i)$ at backend hooks; report
         \textsc{TaskDone}$(e)$ after the final segment. \\
    P2 & After publishing $i$, wait while $i\ge p$.  DeepEP data movement also
         requires receive counts. \\
    D1 & If $p>0$, wait for $r\ge p$ or \textsc{TaskDone}$(e)$; target 0 needs
         no publication.  Enter \textsc{PreDRendezvous}. \\
    D2 & Enqueue the registered D calls. \\
    D3 & After those launch APIs return, enter \textsc{PostDRendezvous}. \\
    D4 & Active ranks set $p\leftarrow p+K$ and release the next P segment.
         Do not wait for P GPU completion. \\
    \bottomrule
  \end{tabular}
  \end{minipage}
\end{center}

Idle ranks issue matched no-work calls, and ranks whose P epoch has completed
continue through the matched control phases.  Neither CPU control phase waits for
CUDA events, stream synchronization, or collective completion; submitted P
work continues.

%% file: supplement/b_lifecycle.tex
\section{Request and KV Lifecycle}
\label{app:lifecycle}

The rank-shared scheduler reserves decode and MTP lookahead capacity before
admitting P.  The following lifecycle rules protect the shared KV storage
while P executes asynchronously.

Each admitted P request reserves KV storage until its prefill result is either
promoted to D or cancelled.  P writes directly into this reservation, and
tensor-parallel workers verify that it still belongs to the request before
promotion.  Results from an earlier request lifetime are discarded;
cancellation waits for submitted P work to drain before the storage is reused.

%% file: supplement/c_waits.tex
\section{Client-Wait Profiles}
\label{app:waits}

For each dependent continuation, the replay driver applies the configured
wait after the preceding turn completes.  P1--P4 scale the same trace-derived
wait vector by 0, 0.01, 0.03, and 0.10, respectively; the unmodified vector is
evaluated separately.  Prompts, output lengths, and dependencies are preserved.

\begin{table}[!b]
  \caption{Configured completion-relative client-wait profiles derived from
  the internal RL trace.}
  \label{tab:pressure-profiles}
  \centering
  \small
  \begin{tabular}{lrrrr}
    \toprule
    Profile & Description & p50 & p90 & p99 \\
    \midrule
    P1 & P-heavy & 0 & 0 & 0 \\
    P2 & short-idle & 31 & 39 & 153 \\
    P3 & mid-idle & 93 & 117 & 459 \\
    P4 & P-light & 310 & 388 & 1,530 \\
    Trace wait & unmodified & 3,104 & 3,884 & 15,296 \\
    \bottomrule
  \end{tabular}
  \vspace{2pt}

  \footnotesize Milliseconds; each system receives the same configured value
  for each of the 1,137 dependent continuations.
\end{table}

%% file: supplement/d_diagnostics.tex
\section{Protocol Diagnostics and Stress Validation}
\label{app:diagnostics}

The crossed-cycle diagnostic uses H200 GPUs; model-free rank-skew stress uses
H20-3E.  These diagnostics are separate from the main performance results.
The controls below distinguish the observed collective-order stall from the
waiting costs of safe coordination.

\begin{table}[t]
  \caption{Crossed-order controls and protocol validation.  The first four
  controls use ATP2/ADP2 except for the single-axis cases.}
  \label{tab:collective-repair}
  \centering
  \small
  \setlength{\tabcolsep}{2.5pt}
  \begin{tabular}{@{}l p{0.56\columnwidth}@{}}
    \toprule
    Case & Outcome \\
    \midrule
    Separate communicators & Recurrent stall \\
    Single-axis topology & Both complete \\
    Residency cap & 23 CTAs/SM: completes; 24: stalls \\
    Selective ordering & 3/3 replays complete \\
    Rank-skew stress & 2,560/2,560 cycles complete \\
    Delayed-P control & D enqueue ACK precedes P completion: 20/64 cycles \\
    \bottomrule
  \end{tabular}
  \vspace{2pt}

  \parbox{\columnwidth}{\footnotesize
  Single-axis: ATP1/ADP4 and ATP4/ADP1.  Rank-skew stress:
  five seeds, up to 20~ms skew.  ACK denotes acknowledgement.}
\end{table}

\subsection{Acknowledgement and Execution Traces}
\label{app:diagnostic-traces}

Global-Enqueue acknowledges all-rank whole-P host submission without a GPU
completion fence.  Fine-grained coordination instead releases P in segments;
Global-Complete waits for whole-P GPU completion.

Each Global-Enqueue epoch uses one rendezvous after 40 dispatch boundaries.
The median across runs of mean acknowledgement latency is 0.93~ms with
trace waits and 0.91~ms without waits.  All six runs acknowledge enqueue before
P GPU completion in some iterations, confirming that this baseline does not
impose a whole-P completion fence.

\textbf{Execution overlap and protocol stress.}
Across a 40-step repaired-path window, each observed rank contains 468 eager P
full-TP NCCL kernels and 57 D graph replays.  The registered NCCL pair has
0~ms overlap.  Segment-level release nevertheless lets D advance
before the complete P epoch: P computation overlaps D for 31.2/31.6~ms on
ranks 0/1 (6.3/6.5\%).

In the matched coordination runs, concurrent P/D GPU work is
90.4/87.6\% under Global-Enqueue and 95.0/94.8\% under Fine-grained for
trace-wait/no-wait workloads, respectively; Global-Complete reaches
89.8/84.1\%.  These measurements describe concurrent P/D GPU work, not
overlap between the ordered NCCL pair.

Table~\ref{tab:collective-repair} also reports
model-free rank-skew stress and a delayed-P control: the former exercises
uneven rank progress, while the latter confirms that D enqueue acknowledgement
does not require whole-P completion.  An eight-rank trace shows matching epoch
and boundary identifiers across ranks.

%% file: supplement/e_segments.tex
\section{Segment Granularity}
\label{app:segments}

The segment policy changes how many safe boundary intervals are released
per round while preserving the same collective order and state isolation.

\begin{figure}[!ht]
\centering
\includegraphics[width=\columnwidth]{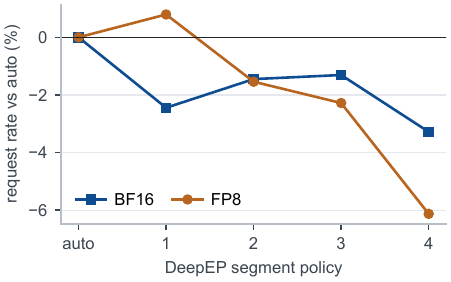}
\caption{Fixed segment spans versus the runtime selector (auto). Request
rate is normalized to auto within each precision.}
\label{fig:segment-policy}
\end{figure}

Figure~\ref{fig:segment-policy} compares fixed segment sizes 1--4 with the
runtime selector.  All six selector runs choose segment 2; that choice is best
for BF16 and within 0.8\% of the best FP8 choice, while a poor fixed choice
loses 6.1\%.  The selected runs have rank-aggregated decode boundary-wait,
prefill wait-for-permit, and prefill enqueue totals within 3.2\% of fixed-2.
From size 1 to 4, aggregate prefill wait-for-permit time falls 31--34\%, while
decode boundary-wait and prefill enqueue time rise 3.9--4.2$\times$ and
3.6--3.7$\times$.  Larger segments reduce P's wait for permission but make D
wait longer for the next boundary.

%% file: supplement/f_live_rl.tex
\section{Live-RL Step-Level Results}
\label{app:live}

\begin{figure}[!ht]
\centering
\includegraphics[width=\columnwidth]{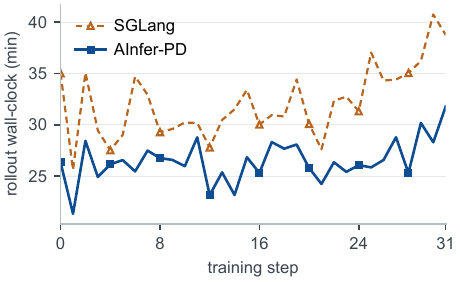}
\caption{Rollout wall-clock time for 32 aligned RL steps in two independent
runs.  \system is lower at every step and by 17.6\% in aggregate.}
\label{fig:live-rl}
\end{figure}

Figure~\ref{fig:live-rl} reports two independent live-RL runs.  Summed
rollout time across steps 0--31 falls from 17.14 to 14.12 hours (17.6\%).
Generated trajectories may diverge after each step, so this is operational
rather than paired evidence.

%% file: references.bib
@inproceedings{zhu2025nanoflow,
  author    = {Kan Zhu and Yufei Gao and Yilong Zhao and Liangyu Zhao and Gefei Zuo and Yile Gu and Dedong Xie and Tian Tang and Qinyu Xu and Zihao Ye and Keisuke Kamahori and Chien-Yu Lin and Ziren Wang and Stephanie Wang and Arvind Krishnamurthy and Baris Kasikci},
  title     = {{NanoFlow}: Towards Optimal Large Language Model Serving Throughput},
  booktitle = {19th USENIX Symposium on Operating Systems Design and Implementation},
  series    = {OSDI '25},
  year      = {2025},
  pages     = {749--765},
  url       = {https://www.usenix.org/conference/osdi25/presentation/zhu-kan}
}

@inproceedings{li2023lina,
  author    = {Jiamin Li and Yimin Jiang and Yibo Zhu and Cong Wang and Hong Xu},
  title     = {Accelerating Distributed {MoE} Training and Inference with {Lina}},
  booktitle = {2023 USENIX Annual Technical Conference},
  series    = {USENIX ATC '23},
  year      = {2023},
  pages     = {945--959},
  url       = {https://www.usenix.org/conference/atc23/presentation/li-jiamin}
}

@inproceedings{zheng2024sglang,
  author    = {Lianmin Zheng and Liangsheng Yin and Zhiqiang Xie and Chuyue Sun and Jeff Huang and Cody Hao Yu and Shiyi Cao and Christos Kozyrakis and Ion Stoica and Joseph E. Gonzalez and Clark Barrett and Ying Sheng},
  title     = {{SGLang}: Efficient Execution of Structured Language Model Programs},
  booktitle = {Advances in Neural Information Processing Systems},
  volume    = {37},
  year      = {2024},
  pages     = {62557--62583},
  url       = {https://papers.nips.cc/paper_files/paper/2024/hash/724be4472168f31ba1c9ac630f15dec8-Abstract-Conference.html}
}

@inproceedings{agrawal2024sarathi,
  author    = {Amey Agrawal and Nitin Kedia and Ashish Panwar and Jayashree Mohan and Nipun Kwatra and Bhargav Gulavani and Alexey Tumanov and Ramachandran Ramjee},
  title     = {Taming Throughput-Latency Tradeoff in {LLM} Inference with {Sarathi-Serve}},
  booktitle = {18th USENIX Symposium on Operating Systems Design and Implementation},
  series    = {OSDI '24},
  year      = {2024},
  pages     = {117--134},
  url       = {https://www.usenix.org/conference/osdi24/presentation/agrawal}
}

@inproceedings{zhong2024distserve,
  author    = {Yinmin Zhong and Shengyu Liu and Junda Chen and Jianbo Hu and Yibo Zhu and Xuanzhe Liu and Xin Jin and Hao Zhang},
  title     = {{DistServe}: Disaggregating Prefill and Decoding for Goodput-Optimized Large Language Model Serving},
  booktitle = {18th USENIX Symposium on Operating Systems Design and Implementation},
  series    = {OSDI '24},
  year      = {2024},
  pages     = {193--210},
  url       = {https://www.usenix.org/conference/osdi24/presentation/zhong-yinmin}
}

@inproceedings{patel2024splitwise,
  author    = {Pratyush Patel and Esha Choukse and Chaojie Zhang and Aashaka Shah and {\'I}{\~n}igo Goiri and Saeed Maleki and Ricardo Bianchini},
  title     = {Splitwise: Efficient Generative {LLM} Inference Using Phase Splitting},
  booktitle = {51st ACM/IEEE Annual International Symposium on Computer Architecture},
  series    = {ISCA '24},
  year      = {2024},
  pages     = {118--132},
  doi       = {10.1109/ISCA59077.2024.00019},
  url       = {https://doi.org/10.1109/ISCA59077.2024.00019}
}

@inproceedings{qin2025mooncake,
  author    = {Ruoyu Qin and Zheming Li and Weiran He and Jialei Cui and Feng Ren and Mingxing Zhang and Yongwei Wu and Weimin Zheng and Xinran Xu},
  title     = {Mooncake: Trading More Storage for Less Computation---A {KVCache-Centric} Architecture for Serving {LLM} Chatbot},
  booktitle = {23rd USENIX Conference on File and Storage Technologies},
  series    = {FAST '25},
  year      = {2025},
  pages     = {155--170},
  url       = {https://www.usenix.org/conference/fast25/presentation/qin}
}

@inproceedings{kamath2025podattention,
  author    = {Aditya K. Kamath and Ramya Prabhu and Jayashree Mohan and Simon Peter and Ramachandran Ramjee and Ashish Panwar},
  title     = {{POD-Attention}: Unlocking Full Prefill-Decode Overlap for Faster {LLM} Inference},
  booktitle = {Proceedings of the 30th ACM International Conference on Architectural Support for Programming Languages and Operating Systems, Volume 2},
  series    = {ASPLOS '25},
  year      = {2025},
  pages     = {897--912},
  doi       = {10.1145/3676641.3715996},
  url       = {https://doi.org/10.1145/3676641.3715996}
}

@inproceedings{feng2025windserve,
  author    = {Jingqi Feng and Yukai Huang and Rui Zhang and Sicheng Liang and Ming Yan and Jie Wu},
  title     = {{WindServe}: Efficient Phase-Disaggregated {LLM} Serving with Stream-Based Dynamic Scheduling},
  booktitle = {Proceedings of the 52nd Annual International Symposium on Computer Architecture},
  series    = {ISCA '25},
  year      = {2025},
  pages     = {1283--1295},
  doi       = {10.1145/3695053.3730999},
  url       = {https://doi.org/10.1145/3695053.3730999}
}

@inproceedings{chen2026muxwise,
  author    = {Yukang Chen and Weihao Cui and Han Zhao and Ziyi Xu and Xiaoze Fan and Xusheng Chen and Yangjie Zhou and Shixuan Sun and Bingsheng He and Quan Chen},
  title     = {Towards High-Goodput {LLM} Serving with Prefill-Decode Multiplexing},
  booktitle = {Proceedings of the 31st ACM International Conference on Architectural Support for Programming Languages and Operating Systems, Volume 2},
  series    = {ASPLOS '26},
  year      = {2026},
  pages     = {2030--2047},
  doi       = {10.1145/3779212.3790236},
  url       = {https://doi.org/10.1145/3779212.3790236}
}

@inproceedings{lin2026bullet,
  author    = {Zejia Lin and Hongxin Xu and Guanyi Chen and Zhiguang Chen and Yutong Lu and Xianwei Zhang},
  title     = {Bullet: Boosting {GPU} Utilization for {LLM} Serving via Dynamic Spatial-Temporal Orchestration},
  booktitle = {Proceedings of the 31st ACM International Conference on Architectural Support for Programming Languages and Operating Systems, Volume 2},
  series    = {ASPLOS '26},
  year      = {2026},
  pages     = {290--306},
  doi       = {10.1145/3779212.3790135},
  url       = {https://doi.org/10.1145/3779212.3790135}
}

@inproceedings{lee2026layeredprefill,
  author    = {Gunjun Lee and Jiwon Kim and Jaiyoung Park and Younjoo Lee and Jung Ho Ahn},
  title     = {From Tokens to Layers: Redefining Stall-Free Scheduling for {MoE} Serving with Layered Prefill},
  booktitle = {Proceedings of Machine Learning and Systems},
  series    = {MLSys '26},
  volume    = {8},
  year      = {2026},
  url       = {https://proceedings.mlsys.org/paper_files/paper/2026/hash/c0f460c6d63599ea870ba9db63dc96a9-Abstract-Conference.html}
}

@misc{hong2025semipd,
  author        = {Ke Hong and Lufang Chen and Zhong Wang and Xiuhong Li and Qiuli Mao and Jianping Ma and Chao Xiong and Guanyu Wu and Buhe Han and Guohao Dai and Yun Liang and Yu Wang},
  title         = {{semi-PD}: Towards Efficient {LLM} Serving via Phase-Wise Disaggregated Computation and Unified Storage},
  year          = {2025},
  eprint        = {2504.19867},
  archiveprefix = {arXiv},
  url           = {https://arxiv.org/abs/2504.19867}
}

@misc{shi2025nexus,
  author        = {Xiaoxiang Shi and Colin Cai and Junjia Du and Zhihao Jia},
  title         = {Nexus: Proactive Intra-{GPU} Disaggregation of Prefill and Decode in {LLM} Serving},
  year          = {2025},
  eprint        = {2507.06608},
  archiveprefix = {arXiv},
  url           = {https://arxiv.org/abs/2507.06608}
}

@inproceedings{sheng2025hybridflow,
  author    = {Guangming Sheng and Chi Zhang and Zilingfeng Ye and Xibin Wu and Wang Zhang and Ru Zhang and Yanghua Peng and Haibin Lin and Chuan Wu},
  title     = {{HybridFlow}: A Flexible and Efficient {RLHF} Framework},
  booktitle = {Proceedings of the Twentieth European Conference on Computer Systems},
  series    = {EuroSys '25},
  year      = {2025},
  pages     = {1279--1297},
  doi       = {10.1145/3689031.3696075},
  url       = {https://doi.org/10.1145/3689031.3696075}
}

@inproceedings{fu2025areal,
  author    = {Wei Fu and Jiaxuan Gao and Xujie Shen and Chen Zhu and Zhiyu Mei and Chuyi He and Shusheng Xu and Guo Wei and Jun Mei and Jiashu Wang and Tongkai Yang and Binhang Yuan and Yi Wu},
  title     = {{AReaL}: A Large-Scale Asynchronous Reinforcement Learning System for Language Reasoning},
  booktitle = {Advances in Neural Information Processing Systems},
  year      = {2025},
  url       = {https://arxiv.org/abs/2505.24298}
}

@misc{hu2026dora,
  author       = {Tianhao Hu and Xiangcheng Liu and Youshao Xiao and Yang Zheng and Xuan Huang and Jinrui Ding and Yufei Zhang and Tao Liang and Hongyu Zang and Quan Chen and Yueqing Sun and Wenjie Shi and Chao Zhang and Wei Wang and Qi Gu and Yerui Sun and Yucheng Xie and Xunliang Cai},
  title        = {{DORA}: A Scalable Asynchronous Reinforcement Learning System for Language Model Training},
  year         = {2026},
  eprint       = {2604.26256},
  archiveprefix= {arXiv},
  url          = {https://arxiv.org/abs/2604.26256}
}

@misc{deepseek2024v3,
  author       = {{DeepSeek-AI}},
  title        = {{DeepSeek-V3} Technical Report},
  year         = {2024},
  eprint       = {2412.19437},
  archiveprefix= {arXiv},
  url          = {https://arxiv.org/abs/2412.19437}
}

@misc{deepseek2025deepep,
  author       = {{DeepSeek-AI}},
  title        = {{DeepEP}: An Efficient Expert-Parallel Communication Library},
  year         = {2025},
  howpublished = {Software artifact},
  url          = {https://github.com/deepseek-ai/DeepEP}
}

@misc{wu2026expertplex,
  author        = {Bingyang Wu and Chao Jin and Zili Zhang and Xinming Wei and Yinmin Zhong and Ruidong Zhu and Chengxu Yang and Xin Jin and Yuliang Liu},
  title         = {{ExpertPlex}: A High-Goodput Disaggregated Serving System for {MoE} {LLM}s with Adaptive Persistent Kernels},
  year          = {2026},
  eprint        = {2607.18002},
  archiveprefix = {arXiv},
  url           = {https://arxiv.org/abs/2607.18002}
}

@inproceedings{pan2025dfccl,
  author    = {Lichen Pan and Juncheng Liu and Yongquan Fu and Jinhui Yuan and Rongkai Zhang and Pengze Li and Zhen Xiao},
  title     = {Comprehensive Deadlock Prevention for {GPU} Collective Communication},
  booktitle = {Proceedings of the Twentieth European Conference on Computer Systems},
  series    = {EuroSys '25},
  year      = {2025},
  doi       = {10.1145/3689031.3717466},
  url       = {https://doi.org/10.1145/3689031.3717466}
}

@inproceedings{gao2026duetserve,
  author    = {Lei Gao and Chaoyi Jiang and Hossein Entezari Zarch and Daniel Wong and Mark D. Hill and Murali Annavaram},
  title     = {{DuetServe}: Harmonizing Prefill and Decode for {LLM} Serving via Adaptive {GPU} Multiplexing},
  booktitle = {Proceedings of the 43rd International Conference on Machine Learning},
  series    = {Proceedings of Machine Learning Research},
  volume    = {306},
  year      = {2026},
  url       = {https://pages.cs.wisc.edu/~markhill/papers/icml2026_DuetServe.pdf}
}

@misc{nvidia2026ncclorder,
  author       = {{NVIDIA Corporation}},
  title        = {{NCCL} User Guide: Using Multiple {NCCL} Communicators Concurrently},
  year         = {2026},
  howpublished = {NVIDIA Documentation},
  url          = {https://docs.nvidia.com/deeplearning/nccl/user-guide/docs/usage/communicators.html#using-multiple-nccl-communicators-concurrently},
  note         = {Accessed: 2026-08-26}
}

@misc{nvidia2026cudastreamorder,
  author       = {{NVIDIA Corporation}},
  title        = {{CUDA} Programming Guide: {CUDA} Stream Ordering},
  year         = {2026},
  howpublished = {NVIDIA Documentation},
  url          = {https://docs.nvidia.com/cuda/cuda-programming-guide/02-basics/asynchronous-execution.html#cuda-stream-ordering},
  note         = {Accessed: 2026-08-26}
}

@misc{nvidia2026cudagraph,
  author       = {{NVIDIA Corporation}},
  title        = {{CUDA} Programming Guide: {CUDA} Graphs},
  year         = {2026},
  howpublished = {NVIDIA Documentation},
  url          = {https://docs.nvidia.com/cuda/cuda-programming-guide/04-special-topics/cuda-graphs.html},
  note         = {Accessed: 2026-08-26}
}
